\documentclass[11pt]{article}
\usepackage{amsfonts}
 \usepackage{amssymb}
 \usepackage{amsmath}
 \def\bsh{\backslash}

 \newfont{\bbbold}{msbm10}

 \def\bbC{\mbox{\bbbold C}}

 \def\cF{{\cal F}}

 \def\cO{{\cal O}}

 \newfont{\goth}{eufm10 scaled \magstep1}

 \def\gu{\mbox{\goth u}}

 \def\a{\alpha}
 \def\b{\beta}
 \def\c{\gamma}
 \def\d{\delta}
 \def\e{\epsilon}
 \def\f{\phi}
 \def\h{\eta}
 \def\k{\kappa}
 \def\L{\Lambda}
 \def\m{\mu}

 \def\s{\sigma}
 \def\t{\tau}
 \def\th{\theta}
 
 \def\adt{\dot \alpha}
 \def\bdt{\dot \beta}
 \def\cdt{\dot\gamma}
 \def\ddt{\dot\delta}
 \def\edt{\dot\epsilon}
 \def\hdt{\dot\eta}
 \def\be{\begin{equation}}\def\ee{\end{equation}}
 \def\bea{\begin{eqnarray}}\def\eea{\end{eqnarray}}
 \def\ba{\begin{array}}\def\ea{\end{array}}

 \def\del{\partial}

 \def\str{\rm str}
 \def\xz{\times}

 \def\nab{\nabla}

 \def\del{\partial}

 \def\3dt{\dot{3}}
 
\def\half{{1\over2}}

 \let\la=\label

 \let\bm=\bibitem{}

 \def\nn{\nonumber}
 \def\bd{\begin{document}}
 \def\ed{\end{document}}
 \def\bea{\begin{eqnarray}}
 \def\ba{\begin{array}}\def\ea{\end{array}}
 \def\eea{\end{eqnarray}}
 \def\ft#1#2{{\textstyle{{\scriptstyle #1}\over {\scriptstyle #2}}}}
 \def\fft#1#2{{#1 \over #2}}
 \newcommand{\eq}[1]{(\ref{#1})}
 \def\eqs#1#2{(\ref{#1}-\ref{#2})}
 \def\det{{\rm det\,}}
 \def\tr{{\rm tr}}\def\Tr{{\rm Tr}}
  \def\str{{\rm str}} \def\diag{{\rm diag}}
 \def\sdet{{\rm sdet}}\def\symtr{{\rm symtr}}

\newcommand{\hoch}[1]{$^{#1}$}

\begin{document}

 \thispagestyle{empty}

 \hfill{KCL-TH-03-02}

  \hfill{\today}

 \vspace{20pt}

 \begin{center}
 {\Large{\bf Integral invariants in $N=4$ SYM and the effective
 action for coincident D-branes}}
 \vspace{30pt}

 {\large J.M. Drummond\hoch1, P.J. Heslop\hoch2, P.S. Howe\hoch3
 and S.F. Kerstan\hoch3}

\vspace{15pt}

\begin{itemize}
\item[$^1$] Department of Mathematics, Trinity College, Dublin,
Ireland. \item[$^2$] II. Institut f\"ur Theoretische Physik,
Universit\"at Hamburg and Institut f\"ur Theoretische Physik,
Universit\"at Leipzig. Germany. \item[$^3$] Department of
Mathematics, King's College, London, U.K.
\end{itemize}

 \vspace{60pt}

 \end{center}

 {\bf Abstract}
The construction of supersymmetric invariant integrals is
discussed in a superspace setting. The formalism is applied to
$D=4,\, N=4$ SYM and used to construct the $F^2$, $F^4$ and $(F^5
+ \del^2 F^4)$ terms in the effective action of coincident
D-branes. The results are in agreement with those obtained by
other methods.  A simple derivation of the abelian $\del^4 F^4$
invariant is given and generalised to the non-abelian case. We
also find some double-trace invariants. The invariants are
interpreted in terms of superconformal multiplets: the $F^2$ and
$F^4$ terms are given by one-half BPS multiplets, the
$(F^5+\del^2F^4)$ arises as a full superspace integral of the
Konishi multiplet $K$ and the abelian $\del^4 F^4$ term comes from
integrating the fourth power of the field strength superfield.
Counterparts of the abelian invariants  are exhibited for the
$D=6,(2,0)$ tensor multiplet and the $D=3, N=8$ scalar multiplet.
The method is also applied to $D=4, N=8$ supergravity. All
invariants in the linearised theory (with $SU(8)$ symmetry) which
arise from partial superspace integrals are constructed.

 {\vfill\leftline{}\vfill \vskip  10pt

 \baselineskip=15pt \pagebreak \setcounter{page}{1}

\section{Introduction}

An intriguing problem in the theory of D-branes is the question of
what is the effective action for a set of coincident branes. In
the abelian case the Born-Infeld approximation,  involving no
derivatives of the field-strength, is well-defined \cite{fradkin},
but it is not clear that there is a meaningful generalisation of
this to the non-abelian case.  In any case, there are also higher
derivative corrections to Born-Infeld for a single brane, so that
it is perhaps more sensible to consider the general effective
action, including all derivative corrections. As yet there is not
a well-established principle for finding this although there have
been some suggestions as to what it might be. The non-abelian
version of the bosonic Born-Infeld action, constructed using the
symmetrised trace, was proposed in \cite{Tseytlin:1997cs}, but it
is known from string theory calculations that this is incomplete
\cite{Kitazawa:xj}. An attempt was made via modified $\k$-symmetry
transformations in \cite{Bergshoeff:2000ik}, but this does not
seem to agree with string theory \cite{Bergshoeff:2001dc}. A
constructive principle, based on the theory admitting solutions of
a particular type has been put forward in \cite{Koerber:2001uu}.
This correctly reproduces the abelian Born-Infeld action
\cite{DeFosse:2001mk} and is so far in agreement with known
results from string theory in the non-abelian case
\cite{Medina:2002nk}. Although it is
difficult to compute the full effective action in a closed form,
calculations so far have been carried out to order $\a'^4$
\cite{Koerber:2002zb}. Further results on derivative corrections
have been obtained in \cite{wyll,bilal}.

The abelian Born-Infeld approximation is also known to be
determined completely by supersymmetry.  The $\k$-symmetric
extension of the Born-Infeld action for a D-brane was discussed in
references \cite{Cederwall:1996pv,
Aganagic:1996pe,Bergshoeff:1996tu}, but perhaps the fact that
supersymmetry determines the action completely is made clearer in
the superembedding formalism \cite{Sorokin:1999jx}. In this
approach there is  a natural constraint on any superembedding
which has a simple geometrical interpretation and which determines
the lowest-order non-linear field equations of most single branes
in type II string theory and M-theory uniquely
\cite{Howe:1996mx,Howe:1996yn}; it also explains the structure of
$\k$-symmetry transformations. Starting from the Wess-Zumino term
one can construct the action, including the Dirac-Born-Infeld term
in the case of D-branes, systematically \cite{hrs}. The
exceptional cases are the branes with low codimension, but here
one can argue, for example in codimension zero
\cite{Drummond:2001uj}, that the multiplet structure of the theory
yields a unique set of constraints which leads to the
$\k$-symmetric Born-Infeld action. However, it is not clear what
the non-abelian generalisation of this is, nor is it clear how
higher-derivative terms in the abelian effective action arise,
although some progress has been made recently \cite{Howe:2001wc}.
In \cite{Drummond:2002kg} a study was made of a toy model which
was designed to represent a set of coincident space-filling branes
in three dimensions. A number of non-abelian extensions of the
abelian super Born-Infeld theory were found all of which
incorporate Tseytlin's action. However, there seems to be no
obvious way of selecting out a unique action. This type of
approach was also advocated for D0-branes in
\cite{Sorokin:2001av} and \cite{Panda:2003dj}

In the superembedding approach for a single brane the target space
supersymmetry is manifest, and one uses the modified field
strength $\cF$ which satisfies the Bianchi identity $d\cF= H$
where $H$ is the pull-back of the NS three-form. This formalism is
not related in a simple way to the usual $N=1,D=10$ superspace
version of the deformed super Maxwell theory, but it is possible
to derive the $D=10, N=1$ superspace constraints corresponding to
the abelian D9-brane action in a systematic manner. This was done
to order $\a'^4$ in \cite{Kerstan:2002au}, but it does not seem to
be straightforward to adapt them to the non-abelian case.

Another way of looking at this problem is provided by spinorial
cohomology. This is a superspace cohomology related to pure
spinors. The relevance of such spinors to supersymmetric field
theories in ten and eleven dimensions was pointed out in
\cite{Howe:mf,Howe:1991bx}. Essentially one considers forms with
only spinorial indices (which are therefore symmetric
multi-spinors) with the gamma-traces removed~\footnote{By this we
mean that, if one contracts on any pair of spinor indices with a
single gamma matrix, one gets zero.}. One then forms a derivative
by acting on such an object with the super-covariant derivative
$D_{\a}$ and projecting out the gamma-trace. In this way one
arrives at a complex and its associated cohomology
\cite{Cederwall:2001dx}. This cohomology is isomorphic to pure
spinor cohomology \cite{Berkovits:2000nn}. Deformations of $D=10$
super Yang-Mills theory are given by elements of the second
spinorial cohomology group with physical coefficients. It is easy
enough to find the first deformation (corresponding to $F^4$)
\cite{brs,Cederwall:2001td} but the analysis at higher orders in
$\a'$ is more difficult, although $\a'^3$ has been studied in the
abelian case \cite{Cederwall:2002df}.

In some recent papers higher-order actions for $D=10$ super
Yang-Mills theory have been constructed, to quadratic order in the
fermions, using supersymmetry (in components) and the  Noether
method. The terms that have been built include the $F^4$ action,
the $F^5$ action, absent in the abelian case, and most recently a
higher derivative abelian $\del^4 F^4$ term
\cite{deRoo:2002ap,Collinucci:2002gd}. The $F^5$ term has also
been discussed in $D=4, N=4$ Yang-Mills theory formulated in $N=1$
superspace \cite{Refolli:2001df,grasso}. These results are in
agreement with those of \cite{Koerber:2001uu}. In
\cite{Koerber:2002zb} the method of \cite{Koerber:2001uu} was used
to construct the purely bosonic terms at $(\a')^4$ in the
non-abelian theory; the result obtained there incorporates the
above $\del^4 F^4$ term in the abelian limit.

In this article we shall rederive these terms in a simple manner
in four dimensions. Four dimensions is easier to work with than
ten because one can make use of harmonic superspace techniques to
construct integrals which involve fewer than the maximum number of
odd coordinates but which are still manifestly
supersymmetric.\footnote{The invariants we find are unique and so
must be the $N=4, D=4$ dimensional reductions of $N=1, D=10$
invariants.} It turns out that the known terms can all be
interpreted in terms of $N=4$ superconformal multiplets. The usual
Yang-Mills action is a component of the supercurrent multiplet,
the $F^4$ term comes from a series C one-half BPS multiplet with
dimension 4 and the $F^5$ term comes from a descendant one-quarter
BPS state. It can alternatively be expressed as a full superspace
integral of the Konishi multiplet, and so is not truly BPS. These
are all of the single-trace terms that can arise as integrals over
fewer than sixteen odd coordinates, although there are also
double-trace one-half BPS and one-quarter BPS terms. There are  no
true series B integrands (which would have to be at least
triple-trace); the only allowed series B integrand is also a
descendant of Konishi and gives the same $F^5$ result. The abelian
$\del^4 F^4$ invariant is a full superspace integral of an
integrand of the form $W^4$ where $W$ is the $N=4$ field strength
superfield whose leading component transforms under the
six-dimensional representation of $SU(4)$. This admits two
single-trace non-abelian generalisations as well as two
double-trace ones. It also generalises to the $D=6, (2,0)$ tensor
multiplet and the $D=3, N=8$ scalar multiplet, these being the
worldvolume multiplets of the M5 and M2 branes respectively. The
sub-superspace integral invariants in these theories, and in $D=4,
N=8$ supergravity (section 5), are examples of BPS contributions
reviewed in detail in \cite{kiritsis}.

The invariants we find are constructed from the on-shell
field-strength superfield. This means that in general they are not
complete. In Appendix B we outline a method for finding these
completions as well as the modifications to the supersymmetry
transformations.

\section{Integral invariants}

The simplest way to construct a supersymmetric integral invariant
is to integrate a superfield over the whole of superspace.
However, it is well-known that one can also obtain invariants by
integrating over a smaller number of $\theta$s; for example, one
can integrate chiral superfields over half of the odd coordinates.
A systematic investigation of $N$-extended $D=4$ Poincar\'e
supersymmetric invariants and the corresponding measures was given
in \cite{Howe:1981xy} where the invariants were dubbed
superactions. The integrands and measures were taken to be Lorentz
scalars, although they are in general not scalars under the
internal $U(N)$ or $SU(N)$ symmetry group. Such measures take a
simpler form in harmonic superspace
\cite{Galperin:1984av,Galperin:uw}; indeed one can reformulate
superactions as harmonic superspace integral invariants where the
integrands are scalars up to internal charges
\cite{Hartwell:1994rp}.

Harmonic superspaces are designed to facilitate the study of
generalised forms of chirality, called Grassmann analyticity, or
G-analyticity for short. G-analytic superfields can be thought of
as depending on fewer odd coordinates than a complete scalar
superfield in ordinary superspace. Harmonic superactions therefore
take the form of integrals of a superfield obeying some particular
G-analyticity constraints (there are several possibilities for
general $N$) matched with the appropriate measure. It should be
noted that, although a generic G-analytic superfield will be a
full superfield in its dependence on all of the reduced set of odd
coordinates, there are a few examples of such superfields which
are ultra-short and which can therefore be integrated over even
fewer odd coordinates than one might have expected at first sight.
This corresponds to the two types of superaction which were
studied in \cite{Howe:1981xy} in terms of the constraints that the
integrands had to satisfy.

The above considerations suggest a way of constructing all
possible invariants in a given theory which involve integrating
over fewer than the maximal number of odd coordinates. If one
knows the multiplets concerned then one can investigate the short
ones, which will usually correspond to G-analytic fields on some
harmonic superspace. In the $N=4$ super Yang-Mills theory, the
field strength superfield is a scalar superfield
$W_{ij},i,j=1\ldots 4,$ transforming under the six-dimensional
representation of $SU(4)$. The simplest multiplets one can
construct are gauge-invariant products of $W$s projected onto
irreducible representations of $SU(4)$. These multiplets are
actually superconformal multiplets and have been widely studied in
the literature. It is therefore a simple task to list the
multiplets and to see which ones can be integrated to give
integral invariants.

There are also superconformal multiplets which are not Lorentz
scalars. However, these can at best be subject to series A-type
shortenings which have the form of spinorial divergences (an
example is the $D=4, N=1$ supercurrent $J_{\a\adt}$ which obeys
the constraints $D^{\a}J_{\a\adt}=\bar D^{\adt}J_{\a\adt}=0$).
However, constrained superfields of this type do not give rise to
integral invariants. In addition, there are multiplets which are
not primary superconformal fields; such fields have leading
components involving the derivatives of the field strength. They
can be expressed as linear combinations of primaries and
descendants of primary fields, or possibly as products of such
fields. In either case it seems highly unlikely that such
multiplets could give rise to integral invariants involving
sub-superspace integrals.

\subsection{Superconformal multiplets}

Representations of $N$-extended superconformal symmetry in $D=4$
are specified by $N+3$ quantum numbers $(L,R,J_1,J_2,a_1,\ldots
a_{N-1})$, where $L$ is the dilation weight, $R$ is the R-charge,
$J_1$ $J_2$ are the two spin quantum numbers, and the $a_i$s are
the Dynkin labels of an irreducible internal $SU(N)$
representation \cite{screp}. The unitary representations have to
satisfy certain unitarity bounds which can be one of three types:

 \be
 \ba{rrclrcl}
{\rm Series\ A}:& L&\geq&2+2J_2-R+{2m \over N}, &L&\geq& 2 +
2J_1+R+2m_1-{2m\over N}\\
&&&&&&\nn\\
{\rm Series\ B}:& L&=&-R +{2m \over N}, &L &\geq&1+m_1 +J_1,
\qquad J_2=0\\
&&&&&&\nn\\
{\rm or}:& L&=&R +2m_1 -{2m \over N}, &L &\geq&1+m_1 +J_2,
\qquad J_1=0\\
&&&&&&\nn\\
{\rm Series\ C}:& L&=&m_1, &R&=&{2m \over N}-m_1, \qquad J_1=J_2=0
\ea\label{abc} \ee

Here $m$ is the total number of boxes in the Young tableau
corresponding to the representation $(a_1,\ldots a_{N-1})$, and
$m_1$ is the number of boxes in the first row. For $N=4$ SYM the
superconformal group is $PSU(2,2|4)$; representations of this
group have $R=0$. For the rest of this section we shall focus on
the $N=4$ case. These representations were discussed in the
context of the AdS/CFT conjecture in \cite{ferrara}.

Series B and C multiplets are always short and there can also be
shortened representations in series A, although these do not
correspond to G-analytic superfields. The series C multiplets can
be either one-half BPS, which means they depend generically on
one-half of the odd coordinates, or one-quarter BPS which depend
essentially on three-quarters of the odd coordinates. Short
multiplets have been discussed in a harmonic superspace framework
in \cite{Hartwell:1994rp,hw,afsz,fs,hesh2}.

The one-half BPS multiplets divide into two cases: the
single-trace operators, also known as CPOs, which correspond to
the supergravity Kaluza-Klein states in the AdS/CFT
correspondence, and multi-trace products of these. The CPO $A_k$
is defined to be the single-trace product of $W^k$ taken in the
representation $[0k0]$ of $SU(4)$. The field strength $W$ ($k=1$)
is only a superconformal field in the free theory and $A_2:=T$ is
the supercurrent multiplet which contains all of the conserved
currents of $N=4$ SYM\footnote{In ordinary superspace the
supercurrent is $\tr(W_{ij}W_{kl}-\frac{1}{24}\e_{ijkl}\e^{mnpq}
W_{mn} W_{pq})$.}. In the interacting theory both $T$ and $A_3$
are extra-short, while any $A_k,\ k\geq 4$ is not subject to any
additional shortening.

The one-quarter BPS multiplets can be subdivided into two classes
as well: the true BPS operators which are protected and which do
not have anomalous dimensions, and those which are descendants of
long operators. In the quantum theory the latter develop anomalous
dimensions and cease to exist as short operators. The true
one-quarter BPS operators are at least double-trace and have
$SU(4)$ Dynkin labels $[pqp]$, while the descendants can be either
single- or multi-trace. A detailed discussion of these operators
and their mixing properties is given in
\cite{Ryzhov:2001bp,D'Hoker:2003vf}.

The series B scalar operators can be thought of as one-eighth BPS
multiplets which depend on seven-eighths of the odd coordinates.
Again there are true multiplets and descendants. The former are at
least triple trace while the latter can be single- or multi-trace.
These operators have $SU(4)$ Dynkin labels $[q+2,p,q]$ (or the
conjugate). The one-quarter BPS multiplets are not subject to
extra shortening, but the one-eighth BPS multiplets which saturate
both unitarity bounds are subject to a second-order spinorial
derivative as well as a first-order G-analyticity constraint.

There are also some series A operators which are shortened. An
example is the square of the supercurrent in the representation
$[020]$. This is subject to a second-order derivative constraint
which survives in the quantum theory. Other series A scalar
operators are long operators which are entire scalar superfields
on ordinary superspace. The simplest example is the Konishi
superfield $K:=\tr(W^2)$ in the singlet representation of $SU(4)$.
In the free theory this superfield satisfies a second-order
constraint  \cite{Howe:1981qj}.

\subsection{Harmonic superspace}

We denote the spinorial derivatives on Minkowski superspace $M_N$
by $(D_{\a i},\bar D_{\bdt}^j), i,j=1\ldots N,$ in two-component
spinor notation. The supersymmetry algebra is $[D_{\a i},\bar
D_{\bdt}^j]=i\d_i{}^j \del_{\a\bdt}$. We define G-analyticity of
type $(p,q)$ to be a set of $p$ $D$s and $q\ \bar D$s which
mutually anti-commute. The space of G-analyticities of type
$(p,q)$ is the coset space $K_{p,q}=S(U(p)\xz U(N-(p+q)\xz
U(q))\bsh SU(N)$. This is easy to see: let $u_I{}^i\in SU(N)$,
where $SU(N)$ acts on $i$ to the right and the isotropy group acts
on $I$ to the left, and let $(u^{-1})_i{}^I$ denote its inverse.
We split the index $I=(r,R,r')$ where $r=1,\ldots p,\ R=p+1\ldots
N-q,\ r'=N-q+1\ldots N$; the isotropy group acts in an obvious
manner. Now set $D_{\a I}:=u_I{}^i D_{\a i}$ and $\bar
D_{\adt}^I:=\bar D_{\adt}^i (u^{-1})_i{}^I$; clearly the
derivatives $(D_{\a r},\bar D_{\adt}^{r'})$ mutually anti-commute.

We define $(N,p,q)$ harmonic superspace to be $M_N\xz K_{p,q}$. A
field on this space is equivalent to a field on $M_N\xz SU(N)$
which is equivariant with respect to the isotropy group, i.e its
dependence on the coordinates of the isotropy group is fixed. Such
a field can be expanded in harmonics on the coset with
coefficients which are conventional superfields whence the
nomenclature. A G-analytic superfield on this space is one which
is annihilated by $(D_{\a r},\bar D_{\adt}^{r'})$; it will
therefore depend on $4N-2(p+q)$ odd coordinates. Fields on
$(N,p,q)$ superspace can also be harmonic analytic which means
they are holomorphic with respect to the $\bar\del$ operator on
$K_{p,q}$; such superfields have finite harmonic expansions since
the coset space is a compact complex manifold, and all of the
supermultiplets which arise in the superconformal context are of
this type.

For $N=4$ SYM we are interested in $(p,q)=(2,2)$ for one-half BPS,
$(p,q)=(1,1)$ for one-quarter BPS and $(p,q)=(1,0)$ for one-eighth
BPS.

\subsubsection*{$(4,2,2)$}

On this superspace we split the index $I=(r,r'),\ r=1,2;r'=3,4$;
the basic G-analytic superfield is

 \be
 W:={1\over2}\e^{rs} u_r{}^i u_s{}^j W_{ij}
 \ee

It is annihilated by $D_{\a r},r=1,2$ and $\bar
D_{\adt}^{r'},r'=3,4$; it is also harmonic analytic on the coset
$S(U(2)\xz U(2))\bsh SU(4)$. The CPOs are given by $A_k=\tr
(W^k)$. The other one-half BPS multiplets are given by products of
the $A_k$s. Note that these superfields depend on half of the odd
coordinates, that is to say, two $\theta$s and two $\bar\theta$s.
Now the highest power of odd variables in $W$ is two, so $T$ has
up to four powers of $\th$ and $A_3$ up to six powers. These
supermultiplets are thus extra short, whereas all the superfields
with $k\geq 4$ depend on all four $\th$s and $\bar\th$s.

\subsubsection*{$(4,1,1)$}

On this space we split the index $I=(1,r,4)$ where $r\in \{2,3\}$.
The basic superfield is

 \be
 W_{1r}:=u_1{}^iu_r{}^j W_{ij}
 \ee

This is easily seen to be $(1,1)$ analytic, i.e. $D_{\a
1}W_{1r}=\bar D_{\adt}^4 W_{1r}=0$. The CPOs are given by single
traces of this superfield with the $SU(2)$ indices symmetrised:

 \be
 A_{r_1\ldots r_k}=\tr (W_{1(r_1}\ldots W_{1 r_k)})
 \ee

The true one-quarter BPS multiplets are given by products of $A$s
with at least one contraction and the remaining $SU(2)$ indices
symmetrised, e.g. the operator $A_{u(rs} T_{t)}{}^u$. An example
of a  descendant is the operator $\tr(Y^2)$ where
$Y:=\e^{rs}W_{1r} W_{1s}$.

\subsubsection*{$(4,1,0)$}

On this space we put $I=(1,r)$ with $r\in\{2,3,4\}$, and define a
field $W_{1r}$ in the same way as in the $(1,1)$ case, although it
now only satisfies the constraint $D_{\a 1}W_{1r}=0$. Again the
CPOs are constructed from single-trace products of this superfield
with all of the $SU(3)$ indices symmetrised. To form a true
one-eighth BPS operator one has to multiply at least three
different $A$s together, contract one (or possibly more) index
from each operator with $\e^{rst}$ and symmetrise on the remaining
indices. The simplest example has dimension six; it is

 \be
 \cO:=\e^{rst} \e^{uvw}T_{ru}  T_{sv}  T_{tw}
 \ee

This operator corresponds to the $SU(4)$ Dynkin labels $[400]$.
The simplest descendant operator in this class is
$\e^{rst}\tr(W_{1r} W_{1s} W_{1t})$ with Dynkin labels $[200]$.

\subsection*{The maximal coset}

Instead of using the coset space $K_{p,q}=S(U(p)\xz U(N-(p+q)\xz
U(q))\bsh SU(N)$ one can replace the middle factor by any subgroup
of $U(N-(p+q))$.  Indeed, in some situations it is useful to
consider the maximal coset space which is obtained by taking the
isotropy group to be the maximal torus, $(U(1))^3$ for $N=4$
\cite{fs,bandos}.  The coset $K:=(U(1))^3\bsh SU(4)$  is the space
of full flags in $\bbC^4$.  In this case we can write an element
of $SU(4)$ as $u_I{}^i=(u_1{}^i,u_2{}^i,u_3{}^i,u_4{}^i)$ and its
inverse by $(u^{-1})_i{}^I$. We have three $U(1)$ charges which we
can take to be $+1$ for each of the indices $1,2,3$. The index $4$
then has charge $-1$ for each $U(1)$. Upper indices have the
opposite charges to lower indices. We can convert $SU(4)$ indices
to $U(1)^3$ indices by means of $u$ and $u^{-1}$, and
differentiation on the coset is carried out using the
right-invariant vector fields $D_I{}^J$. The set $\{D_I{}^J|I<J\}$
corresponds to the components of the $\bar\del$ operator on $K$
while the set $\{D_I{}^J|I>J\}$ corresponds to the components of
the conjugate operator $\del$. The diagonal derivatives are
associated with the isotropy algebra; we can define them in
accordance with the above charge assignments. The three $\gu(1)$
generators are $D^{(r)}:=D_r{}^r,\ r=1,2,3$ where there is no sum,
and we have
 \bea
 D^{(r)} u_s{}^i&=& \d_s{}^r u_s{}^i ,\qquad {\rm no\ sum\ on\ }s\nn\\
 D^{(r)} u_4{}^i&=& -u_4{}^i, \qquad\  r=1,2,3
 \eea
The charges carried by the coset space derivatives are then given
by their numerical indices so that, for example, $D_1{}^2$ has
charge $+1$ with respect to the first $U(1)$and charge $-1$ with
respect to the second.

The properties of the Yang-Mills field strength superfield
$W_{ij}$ are listed in the appendix. In harmonic superspace we
define $W_{IJ}=u_I{}^i u_J{}^j W_{ij}$. The G-analyticity
conditions satisfied by $W$ are
 \bea
 \nab_{\a I} W_{IJ}&=&0,\ \ \ {\rm no\ sum\ on\ }I\nn\\
 \bar \nab_{\adt}^K W_{IJ}&=&0,\ \ \ {\rm if\ } K\neq I,J
 \eea
where $\nab$ denotes the superspace gauge-covariant derivative. In
this superspace one can therefore work with numerical indices but
still retain $SU(4)$ covariance provided that each charge in an
invariant vanishes.

\subsection{Invariants}

To form an invariant we now have to integrate one of the above
supermultiplets over the appropriate measure. There is only one
special case, the supercurrent $T$, which is extra short and
one-half BPS. It depends on essentially four odd coordinates which
suggests that there should be an invariant of the form

 \be
 I_0=\int\, d\m\, T \sim \int\,d^4x\,\left( F^2 + \ldots\right)
 \ee

where the measure is

 \be
 d\m:=d^4x\,du\,[D_3 D_4]^2
 \ee

with $D^2:={1\over2}D_{\a} D^{\a}$ for any $D$, and where $du$
denotes the standard Haar measure on the internal coset $K_{2,2}$.
This expression is not manifestly supersymmetric; the proof that
it is was given in terms of superactions in \cite{Howe:1981xy}.
There is no integral invariant one can form using $A_3$ so all of
the rest are standard harmonic superspace integrals.

The one-half BPS multiplets can be integrated with respect to the
measure

 \be
 d\m_{2,2}:=d^4x\,du\,[D_3 D_4 \bar D^1\bar D^2]^2
 \ee

However, there is a $U(1)$ factor in the isotropy group and the
measure has charge $-4$ with respect to this group. The only
integrands that are allowed will therefore have $U(1)$ charge
$+4$, and there are just two possibilities, $A_4$ and $T^2$. The
first of these, $\int\,d\m_{2,2}\, A_4$ gives the $F^4$ term in
the non-abelian Born-Infeld action while the second gives a
 $(\tr(F^2))^2$ term.

The one-quarter BPS multiplets can be integrated with the measure

 \be
 d\m_{1,1}:=d^4x\,du\,[D_2 D_3 D_4 \bar D^1\bar D^2\bar D^3]^2
 \ee

There is only one single-trace possibility which has the right
charges; it has $SU(4)$ Dynkin labels $[202]$ and can only be
realised by the descendant operator $\tr(Y^2)$ discussed above. In
fact this operator can be written as

 \be
 \tr(Y^2)= [D_1\bar D^4]^2 K
 \ee

where $K:=\tr(W_{ij}\bar W^{ij})$ is the Konishi superfield. We
therefore have

 \be
 I_3=\int\,d\m_{1,1}\tr(Y^2)=\int\, d^4 x\,d^{16}\th K
 \sim \int\,d^4 x \left(F^5 +
 \del^2 F^4 + \ldots\right)
 \ee

There is also a double-trace one-quarter BPS multiplet in the
representation $[202]$; it is given on $(4,1,1)$ superspace by
$T_{rs} T^{rs}$. The integral of this should give rise to
double-trace $\del^2 F^4$ terms.

For the one-eighth BPS supermultiplets the measure is

 \be
 d\m_{1,0}:=d^4x\,du\,[D_2 D_3 D_4 \bar D^1\bar D^2\bar D^3\bar D^4]^2
 \ee

There is only one possible integrand, the single-trace descendant
$\e^{rst}\tr(W_{1r} W_{1s} W_{1t})$. In fact, it can be written as
$(D_1)^2 K$, so that the invariant is just the $F^5$ one given
above.

The above invariants are the only ones that can be constructed in
$N=4$ SYM as integrals involving fewer than sixteen $\th$s. We
note that there is no independent $F^6$ term which confirms that
this term, present in Born-Infeld, is generated by the $F^4$ term,
as shown in \cite{Collinucci:2002gd}. This result has also been 
established directly in four dimensions in the off-shell $N=3$
superspace formalism \cite{Ivanov:2001ec}.

The simplest integrals that can be constructed using the full
superspace measure, apart from the one we have discussed, involve
four powers of $W$. If we switch to $SO(6)$ notation and regard
the real superfield $W_A, A=1,\ldots 6,$ (on Minkowski superspace)
as a vector under this group, we can write the possible integrands
as $\symtr (W_A W_A W_B W_B)$, $\tr([W_A,W_B][W_A,W_B])$, $K^2$
and $T_{AB} T_{AB}$, where $T_{AB}$ denotes the supercurrent which
is a symmetric traceless tensor in this notation. The first of
these is the non-abelian version of the $\del^4 F^4$ term found in
\cite{Collinucci:2002gd}, the second is a similar term which
vanishes in the abelian case and the last two are double-trace
expressions which both reduce to the first integrand in the
abelian case.

\section{F terms}

In this section we evaluate the pure $F$ contributions to these
integrals using the formulae given in the appendix. It is
convenient to use the maximal coset superspace for this. The
$\a'^2$ term is well-known, so we shall not give it again here; it
is easy to check that it does indeed have the correct Born-Infeld
form. We shall also restrict our attention to single-trace
integrands.

\subsection*{$\cO(\a'^3)$}

The invariant can be written, using the maximal coset, as

\be I_3=\int d^4 x\,du\, d^{16}\th\, \tr (W_{12} W_{34}) \ee

This expression can easily be seen to be the same as the integral
over ordinary superspace of $\tr(W_AW_A)$. However, it is useful
to use harmonic notation even for full superspace integrals as it
is easier to evaluate them this way.

As usual, we can carry out the integration over the odd
coordinates by applying eight $D$s and $\bar D$s to the integrand.
We can substitute these by gauge-covariant  spinorial derivatives,
as the integrand is gauge-invariant, and so the task is to
evaluate

\be \tr\, [\nab_1 \nab_2 \nab_3 \nab_4 \bar{\nab}^4 \bar{\nab}^3
  \bar{\nab}^2 \bar{\nab}^1]^2 (W_{12} W_{34}),
\ee

in terms of the component fields. The pure field strength
contribution is

\begin{align}
I_3(F)= \tr \bigl( & 6 M_{\a}{}^{\b}M_{\b}{}^{\c}M_{\c}{}^{\a}
\bar M_{\adt}{}^{\bdt} \bar M_{\bdt}{}^{\adt} -2
M_{\a}{}^{\b}M_{\b}{}^{\c} \bar M_{\adt}{}^{\bdt} M_{\c}{}^{\a}
\bar M_{\bdt}{}^{\adt}\nn \\
- & 6 \bar M_{\adt}{}^{\bdt} \bar M_{\bdt}{}^{\cdt} \bar
M_{\cdt}{}^{\adt} M_{\a}{}^{\b} M_{\b}{}^{\a} + 2 \bar
M_{\adt}{}^{\bdt} \bar M_{\bdt}{}^{\cdt} M_{\a}{}^{\b}
\bar M_{\cdt}{}^{\adt} M_{\b}{}^{\a}\nn \\
+ & \bar M_{\adt}{}^{\bdt} \bar M_{\bdt}{}^{\adt} \nab_{\c \cdt}
M_{\a}{}^{\b} \nab^{\c \cdt} M_{\b}{}^{\a} + M_{\a}{}^{\b} \bar
M_{\adt}{}^{\bdt} \nab_{\c \cdt} M_{\b}{}^{\a}
\nab^{\c \cdt} \bar M_{\bdt}{}^{\adt}\nn \\
+ & M_{\a}{}^{\b} \bar M_{\adt}{}^{\bdt} \nab_{\c \cdt} \bar
M_{\bdt}{}^{\adt} \nab^{\c \cdt} M_{\b}{}^{\a} + \bar
M_{\adt}{}^{\bdt} M_{\a}{}^{\b} \nab_{\c \cdt} M_{\b}{}^{\a}
\nab^{\c \cdt} \bar M_{\bdt}{}^{\adt}\nn \\
+ & \bar M_{\adt}{}^{\bdt} M_{\a}{}^{\b} \nab_{\c \cdt} \bar
M_{\bdt}{}^{\adt} \nab^{\c \cdt} M_{\b}{}^{\a} + M_{\a}{}^{\b}
M_{\b}{}^{\a} \nab_{\c \cdt} \bar M_{\adt}{}^{\bdt} \nab^{\c \cdt}
M_{\bdt}{}^{\adt} \bigr)
\end{align}

where $M_{\a\b}$ is the field strength tensor in spinor notation,
$F_{\a\adt,\b\bdt}=\e_{\adt\bdt} M_{\a\b}-\e_{\a\b}\bar
M_{\adt\bdt}$. This result agrees with other calculations reported
in the literature.

It is easy to see that this vanishes in the abelian limit. In this
case, the Konishi multiplet obeys the constraint $D_{ij} K=0,\
D_{ij}:=\frac{1}{2}\e^{\a\b} D_{\a i} D_{\b j}$, and so we have a
full superspace integral of a constrained superfield which is
trivially zero.

\subsection*{$\cO(\a'^4)$}

At $\a'^4$, in the abelian case, there is an invariant of the form

\be I_4=\int d^4 x\, d^{16}\th\, (W_A W_A)^2 =\int d^4 x\,du\,
d^{16}\th\, W_{12}^2 W_{34}^2. \ee

It is simple to evaluate the pure field strength contribution to
this invariant. We find

\begin{align}
I_4(F)=&M_{\a \b} M^{\a \b} \del_{\e \edt} \del_{\h \hdt} \bar
M_{\adt \bdt}
\del^{\e \edt} \del^{\h \hdt} M^{\adt \bdt} \notag \\
+4 &M_{\a \b} \del_{\e \edt} \del_{\h \hdt} M^{\a \b} \bar M_{\adt
\bdt}
\del^{\e \edt} \del^{\h \hdt} M^{\adt \bdt} \notag \\
+ &\del_{\e \edt} \del_{\h \hdt} M_{\a \b} \del^{\e \edt} \del^{\h
\hdt} M^{\a \b} \bar M_{\adt \bdt} \bar M^{\adt \bdt}.
\label{a4abelian}
\end{align}

This result should be compared with that of
\cite{Collinucci:2002gd}, where higher order supersymmetric
actions for the $N=1$, $D=10$ Maxwell multiplet were computed
using the  Noether procedure. The $\a'^4$ terms computed there
fall into two groups. The first group consists of the terms that
are required to continue the Born-Infeld invariant up to $\a'^4$
and these are induced by the corrections at $\a'^2$. The second
group contains terms of the form $\del^4 F^4$ and represent the
start of a new, independent, invariant. The result \eq{a4abelian},
which is necessarily the start of a new invariant, agrees with the
dimensional reduction of the second group of $\a'^4$ terms found
in \cite{Collinucci:2002gd} to four dimensions.

This invariant can be generalised immediately to the non-abelian
case. The simplest single-trace invariant is

 \be
 I_4=\int d^4 x\,  d^{16} \th\, \symtr(W_A W_A W_B W_B)=\int
 d^4 x\, du\, d^{16} \th\, \symtr (W_{12} W_{12} W_{34} W_{34}).
 \ee

To calculate the component description one then has to perform the
differentiation. The task is to evaluate

\be \symtr [\nab_1 \nab_2 \nab_3 \nab_4 \bar{\nab}^4 \bar{\nab}^3
\bar{\nab}^2 \bar{\nab}^1]^2 (W_{12} W_{12} W_{34} W_{34}). \ee

This is straightforward, if a little tedious. The pure field
strength contribution is

\begin{align}
I_4(F)= \symtr \Bigl( -2 \bar M_{\cdt \ddt} \bar M^{\cdt \ddt}
\bigl( & 2 [[M_{\c \d},M_{\b}{}^{\c}],M^{\d}{}_{\a}]M^{\a \b}
-[\nab_{\b \bdt} M_{\c \d},\nab_{\a}{}^{\bdt} M^{\c \d}]M^{\a \b}
\notag \\
&-5[\nab_{\b \bdt} M_{\c \d},M^{\d}{}_{\a}]\nab^{\a \bdt} M^{\b
\c}
+ [M_{\b \d},M_{\a}{}^{\d}][M_{\c}{}^{\b},M^{\c \a}]  \notag \\
&- 2[M_{\b \d},M_{\a \c}][M^{\c \d},M^{\a \b}]
+\frac{1}{2}\nab_{\a \adt}\nab_{\b \bdt}M_{\c \d} \nab^{\a
\adt}\nab^{\b \bdt}M^{\c \d} \bigr)
\notag \\
-4 \bar M_{\ddt}{}^{\cdt}M_{\c \d} \bigl( &-2[[\bar
M^{\ddt}{}_{\cdt},M_{\b}{}^{\c}],M_{\a}{}^{\d}]M^{\a \b} +
2[\nab_{\b \bdt} \bar M^{\ddt}{}_{\cdt},\nab_{\a}{}^{\bdt}
M^{\c \d}]M^{\a \b} \notag \\
& +4[\nab_{\b \bdt} \bar M^{\ddt}{}_{\cdt},M_{\a}{}^{\d}] \nab^{\a
\bdt}M^{\b \c} +[M_{\b}{}^{\d},M_{\a}{}^{\c}][\bar
M^{\ddt}{}_{\cdt},M^{\a \b}]
\notag \\
&+[\bar M^{\ddt}{}_{\cdt},\nab_{\a \adt}M_{\b}{}^{\c}] \nab^{\a
\adt}M^{\b \d}
 - 2[\bar M^{\ddt}{}_{\bdt},M_{\b}{}^{\d}][M^{\b \c},
\bar M_{\cdt}{}^{\bdt}]\notag \\
&-\frac{1}{2}\nab_{\a \adt} \nab_{\b \bdt} \bar M^{\ddt}{}_{\cdt}
\nab^{\a \adt} \nab^{\b \bdt} M^{\c \d} \bigr) \notag \\
+2 M_{\c \d} M^{\c \d} &[\nab_{\b \bdt}\bar M_{\cdt
\ddt},\nab_{\a}{}^{\bdt} \bar M^{\cdt \ddt}]M^{\a\b} \Bigr)+
\text{ conjugate}
\end{align}

There is a second invariant which vanishes in the abelian limit;
it is

 \be
 I'_4=\int d^4 x\,d^{16}\th\,\tr([W_A,W_B] [W_A,W_B])=\int
 d^4 x\, du\, d^{16} \th\, \tr ([W_{12}, W_{34}] [W_{12},W_{34}])
 \ee

We believe that $I_4$ and $I'_4$ should correspond to the terms
$L_{4,2}$ and $L_{4,4}$ given in \cite{Koerber:2002zb}, although
we have not checked this in detail. However, $N=4$ supersymmetry
by itself does not fix the relative coefficient between the two
terms, in contrast to the approach of \cite{Koerber:2002zb}.

At higher powers in $\a'$ there will be more and more terms that
can be written down. We shall not attempt to classify these. There
are, however, only a few which arise from powers of $W$ which give
rise to pure $F$ component Lagrangians. We briefly discuss two of
these.

\subsection*{$\cO(\a'^5)$}

The simplest invariant at this order is

  \be
  I_5=\int d^4 x\, du\,  d^{16} \th\, \symtr (W_{12}^3 W_{34}^3).
  \ee

After a little work the pure field strength contribution can be
extracted from this. It is

\begin{align}
I_5(F)=\symtr(& \bar M_{\cdt \ddt} \bar M^{\cdt \ddt} M_{\c \d}
\bar M_{\adt \bdt} \bar
M^{\adt \bdt} [M^{\d}{}_{\b},M^{\b \c}]\nn\\
- &M_{\c \d} M^{\c \d} \bar M_{\cdt \ddt} M_{\a \b} M^{\a \b}
[\bar M^{\ddt}{}_{\bdt},\bar M^{\bdt \cdt}]\nn\\
+ & 8 M^{\a \b} \bar M^{\adt \bdt} \nab_{\a \adt} M_{\c \d} M^{\c
\d} \nab_{\b \bdt} \bar M_{\cdt \ddt} \bar M^{\cdt \ddt})
\end{align}

We observe that this does not vanish in the abelian limit and so
provides an example of an invariant of the form $\del^2 F^6$. Such
terms are claimed to absent in the abelian effective action
\cite{Collinucci:2002gd}, but this one seems to be allowed by
four-dimensional $N=4$ supersymmetry.

\subsection*{$\cO(\a'^6)$}

At order $\a'^6$ there is an invariant of the form

 \be
 I_6=\int d^4x\,du\,  d^{16} \th\, \symtr (W_{12}^4 W_{34}^4)
  \ee

which has as its pure field strength part

 \be
 I_6(F)=\symtr (M_{\a \b} M^{\a \b} \bar M_{\adt \bdt} \bar M^{\adt
 \bdt})^2. \la{f8}\ee

Note that this $F^8$ term is not the same as the Born-Infeld $F^8$
contribution. The latter is

\be \frac{1}{2}\left(M^2\right)^3\bar{M}^2+\frac{15}{8}
M^2M^2\bar{M}^2\bar{M}^2-\frac{1}{2}M^2\left(\bar{M}^2\right)^3
\ee

At orders beyond $\a'^6$ there are no invariants constructed just
from integrals of products of $W$s which contribute to the pure
field strength part of the action.

\section{Other models with sixteen supersymmetries}

The formalism described above can easily be applied to other
models with sixteen supersymmetries such the $D=3, N=8$ scalar
multiplet, the worldvolume multiplet of the M2-brane, and the
$D=6,(2,0)$ tensor multiplet, the worldvolume multiplet of the
M5-brane. In these cases we do not know what the non-abelian
theories are but we can write down some abelian invariants. Both
multiplets can be described by  Lorentz scalar analytic
superfields $W$ on appropriate harmonic superspaces
\cite{Howe:1998jw}. These superspaces are a little more
complicated to describe than the four-dimensional ones because the
internal symmetry groups are orthogonal rather than unitary;
details of these superpaces and superconformal fields on them can
be found in the literature
\cite{Leeming,Howe:1998jw,Ferrara:2000xg,Ferrara:2000ki}.

As in the $N=4$ Maxwell case there are only two invariants which
can be constructed by integrating over fewer than sixteen odd
coordinates. These are, schematically, $d^4\th\, W^2$ and
$d^8\th\, W^4$. The first of these corresponds to the linearised
on-shell action, while the second is the four-field contribution
to the brane action which is of generalised Born-Infeld type for
the five-brane. The quadratic terms vanish on-shell at the
linearised level.  These BPS type terms, together with
higher-order terms which are required by consistency with
supersymmetry, will give rise to the known dynamics of these
branes. In fact, for the 5-brane, there is no Lorentz covariant
action involving just the fields of the multiplet due to the
self-duality of the three-form field strength tensor\footnote{Such
an action can be written with the aid of an additional scalar
field and gauge invariance \cite{Pasti:1997gx}.}. This holds for
the non-linear Born-Infeld type theory, but it seems that it is
only the $F^2$ term that has this problem if we try to expand the
action in powers of $F$.

For both multiplets the first non-trivial corrections to the known
brane dynamics are therefore given, at the linearised level, by
integrals of the form $d^{16}\th W^4$. For the five-brane this
again gives a term of the form $\del^4 F^4+\ldots$, where $F$ is
the three-form field strength of the tensor multiplet, while in
the membrane case the bosonic part of the invariant is a quartic
expression in the extrinsic curvature. A complete analysis of the
latter in the non-linear theory is in preparation \cite{hklt}.

\section{$N=8$ invariants}

In this section we briefly discuss the integral invariants in
$N=8$ supergravity that can be constructed form the linearised
field strength $W_{ijkl},\ i=1\ldots 8,$ which are $SU(8)$
invariant and which are integrals over fewer than thirty-two odd
coordinates. The superfield $W_{ijkl}$ is totally antisymmetric
and transforms under the seventy-dimensional real representation
of $SU(8)$. It obeys the constraints

 \bea
 \bar W^{ijkl}&=&\frac{1}{4!}\e^{ijklmnpq}W_{mnpq}\\
 D_{\a i}W_{jklm}&=&D_{\a [i}W_{jklm]}\\
 \bar D_{\adt}^i W_{jklm}&=&-\frac{4}{5}\d^i_{[j}\bar D^n_{\adt} W_{klm]n}
 \eea

the third of which follows from the other two. This superfield
defines an ultra-short superconformal multiplet. It can be
represented on various harmonic superspaces in a similar manner to
the $N=4$ SYM field strength superfield.

We can again construct superconformal multiplets by taking
products of $W$ projected into irreducible representations of
$SU(8)$. Although the $N=8$ superconformal group has an R
generator, these representations have $R=0$ because $W$ itself
carries no R charge. It turns out that the only series C BPS
multiplets that give rise to invariant integrals can be written on
$(8,p,p)$ superspace with $p\leq 4$. The only non-zero quantum
numbers are the $SU(8)$ Dynkin labels $a_p=a_{8-p}$, $a_4$ and the
dilation weight $L=2a_p$. Moreover, because of the structure of
the measure, all of the integrands have $L=4$; only for $p=4$ is
$a_4\neq 0$.

\subsection*{$(8,4,4)$}

The index $I$ is split into two blocks of four; the superfield can
be written

 \be
 W_{1234}:=u_1{}^i u_2{}^j u_3{}^k u_4{}^l  W_{ijkl}
 \ee

It is a singlet under both $SU(4)$s. The invariant is

 \be
 I_3=\int\ d\m_{4,4} (W_{1234})^4
 \ee

where

 \be
 d\m_{4,4}:=d^4x\,du\, [D_5 D_6 D_7 D_8\bar D^1 \bar D^2 \bar
D^3\bar D^4]^2
 \ee

This is the well-known three-loop counterterm
\cite{Kallosh:1980fi,Howe:1981xy,Hartwell:1994rp}. If we carry out
the odd integrations we find the supersymmetric completion of the
square of the Bel-Robinson tensor.

\subsection*{$(8,3,3)$}

In this superspace we split $I=(r,R,r')$ where
$r\in\{1,2,3\};\,r'\in\{6,7,8\};\, R\in \{4,5\}$. The superfield
is

 \be
 W_{123R}:=u_1{}^i u_2{}^j u_3{}^k u_R{}^l W_{ijkl}
 \ee

The $SU(8)$ representation is $[0020200]$ and the integrand should
be of the form $W^4$. However, it is easy to see that no such term
can be constructed which is invariant under the central $SU(2)$.

\subsection*{$(8,2,2)$}

In this case the central index $R\in\{3,4,5,6\}$ and the
superfield is

 \be
 W_{12RS}:=u_1{}^i u_2{}^j u_R{}^k u_S{}^l  W_{ijkl}
 \ee

In this case we can form the invariant

 \be
 I_5=\int\ d\m_{2,2} (\e^{RSTU}W_{12RS} W_{12TU})^2\sim
 \int\,d^4x\, \del^4 R^4
 \ee

where

 \be
 d\m_{2,2}:=d^4x\,du\, [D_3 D_4 D_5 D_6 D_7 D_8\bar D^1 \bar D^2 \bar
D^3\bar D^4\bar D^5\bar D^6]^2
 \ee

\subsection*{$(8,1,1)$}

In this space the central index $R$ runs from $2$ to $6$ and the
superfield is

 \be
 W_{1RST}:=u_1{}^i u_R{}^j u_S{}^k u_T{}^l W_{ijkl}
 \ee

The invariant is

 \be
 I_6=\int\ d\m_{1,1} X_R{}^S X_S{}^R\sim
 \int\,d^4x\, \del^6 R^4
 \ee

where

 \be
 X_R{}^S:= \e^{ST_1\ldots T_5} W_{RT_1 T_2} W_{T_3 T_4 T_5}
 \ee

and where the measure is defined in the obvious way.

These are all the invariants that can be constructed from series C
BPS multiplets, but there is also a series B candidate. These
multiplets can be realised on $(8,p,0)$ superspace, where $p\leq
4$. In order to be scalars under the internal group $SU(p)$ they
can only have $a_p\neq 0$. The only possibility has $p=2$ and
$L=3$. The superfield is $W_{12rs}$ where $r,s\in\{3,\ldots 8\}$
and the putative invariant is

 \be
 I=\int\, d\m_{2,0}\, \e^{r_1\ldots r_6} W_{12r_1r_2}  W_{12r_3r_4}
 W_{12r_5r_6}
 \ee

However, the integral turns out to vanish.  This is easy to see
from representation theory. The multiplet given by the integrand
satisfies both of the series B unitarity bounds. This means that
it satisfies a first-order G-analyticity $D$-constraint and a
second-order $\bar D$ constraint. Since we are integrating over
all of the $\bar\th$s it therefore follows that we must get zero.

There are therefore just three $N=8$ superinvariants which involve
sub-superspace integrals. In the context of quantum supergravity
they can be interpreted as possible four-point counterterms at
three, five and six loop order respectively. It has been argued
that the coefficient of the three-loop counterterm vanishes, and
there is also a question mark concerning the coefficient of the
five-loop counterterm \cite{Bern:2000mf}. In \cite{Howe:2002ui} it
was pointed out that the vanishing of these coefficients can
probably be explained be the existence of an off-shell version of
$N=8$ supergravity with more than half of the supersymmetries made
manifest. If one can quantise with manifest $N=6$ supersymmetry,
then the onset of divergences would be expected to arise at five
loops, while if one can maintain $N=7$ supersymmetry it should
occur at six loops.

These invariants can also be interpreted as dimensional reductions
of higher-order terms in the effective actions of type II string
theories or M-theory.

\section{Conclusions}

In this paper we have used superspace methods to construct
integral invariants in super Yang-Mills theory and linearised
supergravity. These are manifestly supersymmetric in terms of the
original on-shell supersymmetry, but the non-linearities which
arise as a consequence of including higher-order terms in the
action can be computed systematically, at least in principle.

In the Yang-Mills case we have reproduced the string tree-level
terms up to order $\a'^4$ which have been found by other means.
The terms up to order $\a'^3$ and the abelian $\a'^4$ term are in
agreement with other calculations. The non-abelian $\a'^4$ terms
seem to be in agreement with those of \cite{Koerber:2002zb}
although we have not checked this in full detail. In addition we
have two independent terms whereas the authors of this paper find
definite relative coefficients.  We have not presented any
fermionic terms in this paper, but it should be straightforward,
if tedious, to construct the complete component actions from our
results.

In the abelian theory it seems possible that the Born-Infeld terms
could be generated by the original action and the first  $F^4$
deformation by the linear supersymmetry.  For example, as we have
seen, there are no independent invariants corresponding to the
Born-Infeld $F^6$ and $F^8$ terms, although there is a different
$F^8$ invariant. In addition, the relative coefficient between the
$F^2$ term and the rest can be adjusted by rescaling $F$. If this
were to be true we could regard the supersymmetric BI action as
the BPS part of the full effective action for a single brane. It
would be tempting to try to extend this definition to the
non-abelian case: the non-abelian Born-Infeld action should then
be the action generated by the two single-trace BPS contributions,
$F^2$ and $F^4$. However, it would be more difficult to  define
the higher-order terms unambiguously in this case because the
rejection of terms other than slowly-varying ones is no longer
valid. It is possible that a criterion for accepting higher-order
terms could be devised using group-theoretical structures, which
could lead, for example, to the identification of the non-abelian
BI sector of the theory with the supersymmetric extension of the
Tseytlin symmetrised trace action. Another possibility is that the
second ($D=10$) supersymmetry could play a more significant
r\^ole. Apart from the BPS invariants the other terms involving
higher derivatives or commutators are, as we have seen, associated
with long multiplets. Without further input the method we have
used here would seem to allow a proliferation of such terms at
higher order in $\a'$. It is possible that some simplifications
might be obtained if we employed the second supersymmetry,
although this is something we have not attempted to do here.

We remark that our interpretation of our results is that they
should be thought of as arising by dimensional reduction from the
effective action of  D9-branes  or directly from the effective
action of D3-branes, in both cases in flat backgrounds. The fact
that our results come out rather naturally in terms of
superconformal multiplets suggests that it might be useful to
consider D3-branes in an $AdS_5\xz S^5$ background, in which case
one would expect to have $N=4$ superconformal symmetry of the
action itself.

Although we have not taken all possible symmetries into account it
is still possible for us to make a few remarks about some of the
conjectures and observations made in \cite{Collinucci:2002gd}. For
example, the fact that the three-point function of the open string
vanishes can easily be understood in superspace. There are no
sub-integrals of this form, and one cannot form a Lorentz and
$SU(4)$ invariant from any combination of three fields in the
Maxwell multiplet. On the other hand, it is possible to construct
odd-point invariants and invariants which do not have any purely
bosonic contributions. This can be done directly in $D=10$
superspace where the superfield $\L^\a$ obeys $D\L\sim F$. An
example of both of these types of behaviour is given by
$\int\,d^{10}x\,d^{16}\th\,\L^{16}(\del_a\L \c_c \del_b\L)\del^c
F^{ab}$. This has nineteen fields and gives rise to a spacetime
invariant in which each term has at least two fermions. However,
this does not mean that such terms should arise in practice. For
example, it may well be the case that all the integrands are given
by superconformal multiplets (in four dimensions), and that, in
the abelian theory at least, they could be invariant under an
additional ``bonus'' $U(1)_Y$ symmetry inherited from IIB
supergravity \cite{Intriligator:1998ig}. If true, this might rule
out the type of terms we have just mentioned.

We have also seen in the non-abelian case that there are
sub-superspace invariants involving double traces which should
correspond to string contributions starting at one loop. There are
just two such BPS terms, one at $\a'^2$ and one at $\a'^3$, and
there are no other BPS integral invariants at higher loops.

\section*{Acknowledgements}

This work was supported in part by EU contracts HPRN-2000-00122
and HPRN-CT-2000-00148 and PPARC grants PPA/G/S/1998/00613 and
PPA/G/O/2000/00451. SFK thanks the German National Merit
Foundation for financial support.

We would like to thank K. Stelle, D. Tsimpis, A. Sevrin
and P. Koerber for stimulating discussions.


\section*{Appendix A: $N=4$ SYM in superspace}

We summarise here our conventions for $N=4$ SYM in superspace
\cite{sohnius}. We use a time-favoured metric and convert from
vector indices to two-component spinor indices by means of the
sigma matrices, e.g. $v_{\a\adt}=(\s^a)_{\a\adt}v_a$ where
$(\s^a)_{\a\adt}=(\sqrt{2})^{-1}(1,i\t^i)$ where $\t^i$ denotes
the usual Pauli matrices. The square root factor means that $u^a
v_a=u^{\a\adt} v_{\a\adt}$, or, equivalently,
$\h_{\a\adt,\b\bdt}=\e_{\a\b}\e_{\adt\bdt}$. Our convention for
the epsilon tensors is that they are all the same numerically,
e.g. $\e_{12}=1$, so that $\e_{\a\b}\e^{\a\c}=\d_{\b}{}^{\c}$.

We suppose the gauge group is $SU(n)$. A $p$-form $\f$ in the
adjoint representation transforms according to the rule

 \be
 \f\mapsto g \f g^{-1},\qquad {\rm where}\ g\in G
 \ee

while the connection $A$ transforms as

 \be
 A\mapsto g A g^{-1} + dg g^{-1}
 \ee

The covariant exterior derivative $D$ acts on $\f$ by

 \be
 D\f=d\f-(-1)^pA\f + \f A
 \ee

from which

 \be
 D^2\f=[\f,F], \qquad {\rm where}\ F=dA+ A^2
 \ee

or, in indices, for a scalar $\f$

 \be
 [\nab_A,\nab_B]\f=-t_{AB}{}^C\nab_C\f +[\f, F_{AB}]
 \ee

where $t_{AB}{}^C$ is the flat torsion which is zero except for
$t_{\a i\bdt}^{\phantom{\a i}j \,c}=-i\d_i{}^j(\s^c)_{\a\bdt}$.
$F$ is Lie algebra valued, and so is skew-hermitian $F=-F^*$.

The constraints on $F_{AB}$ are

 \be
 F_{\a i\b j}=\e_{\a\b}W_{ij};\qquad F_{\a i \bdt}^{\phantom{\a i}j}=0
 \ee

which implies that $F_{\adt\bdt}^{ij}=\e_{\adt\bdt}\bar W^{ij}$,
where the bar denotes hermitian conjugation. We also impose the
self-duality constraint

 \be
 \bar W^{ij}=\half\e^{ijkl} W_{kl}
 \ee

One now employs the Bianchi Identities to find the other
components of $F_{AB}$ and the contents of the superfield
$W_{ij}$. We find

 \bea
 F_{\a i,\b\bdt}&=&-i\e_{\a\b}\bar \L_{\bdt i}\qquad
 F_{\adt,\b\bdt}^i=-i\e_{\adt\bdt} \L^i_{\b}\nn\\
 &&\nn\\
 F_{\a\adt,\b\bdt}&=& \e_{\adt\bdt}M_{\a\b}-\e_{\a\b}\bar
 M_{\adt\bdt}
 \eea

where the second equation expresses the spacetime field strength
in terms of a symmetric bi-spinor. We then have the following
relations for the derivatives of the superfield $W_{ij}$,

 \begin{alignat}{3}
 \nab_{\a i} W_{jk}&=\e_{ijkl}\L^l_\a \hspace{4cm}&\bar\nab_{\adt}^i\bar
 W^{jk}&=\e^{ijkl}\bar \L_{\adt l}\nn \\
 &&&\nn\\
 \bar \nab_{\adt}^i W_{jk}&=2\d^i_{[j}\bar \L_{\adt k]}
 &\nab_{\a i}\bar W^{jk}&=2\d_i^{[j}\L_\a^{k]}\nn\\
 &&&\nn\\
 \nab_{\a i}\bar\L_{\bdt j}&=i\nab_{\a\bdt}
 W_{ij}&\bar\nab_{\adt}^i\L_\b^j&=i\nab_{\b\adt}\bar W^{ij}\nn\\
 &&&\nn\\
 \nab_{(\a
 i}\L_{\b)}^j&=-\d_i^jM_{\a\b}&\bar\nab^i_{(\adt}\bar\L_{\bdt)j}&=\d^i_j\bar
 M_{\adt\bdt}\nn\\
 &&&\nn\\
 \nab_{\a i}\L^{\a j}&=[W_{ik},\bar W^{jk}] &\nab_{\a
 i}\nab_j^\a\bar W^{kl}&=2\d_j^{[k}[W_{im},\bar W^{l]m}]\nn\\
 &&&\nn\\
 \bar\nab_{\adt}^{i}\bar\L^{\adt}_{j}&=-[W_{jk},\bar
 W^{ik}]&\bar\nab_{\adt}^{i}\bar\nab^{\adt j} W_{kl}&=
 -2\d^j_{[k}[W_{l]m},\bar W^{im}]\nn\\
 &&&\nn\\
 \nab_{\a i} M_{\b\c}&=-i\e_{\a(\b}\nab_{\c)\cdt}\bar\L_i^{\cdt}&
 \bar\nab^i_{\adt}\bar
 M_{\bdt\cdt}&=i\e_{\adt(\bdt}\nab_{\cdt)\c}\L^{\c i}\nn\\
 &&&\nn\\
 \bar\nab_{\adt}^i M_{\b\c}&=-i\nab_{\adt(\b}\L^i_{\c)}&
 \nab_{\a i}\bar M_{\bdt\cdt}&=i\nab_{\a(\bdt}\bar\L_{\cdt)i}
 \end{alignat}

From these relations it is clear that the only independent
component fields in $W_{ij}$ are the physical fields of the $N=4$
SYM multiplet, and so they must obey their equations of motion.
For example, the spinor equation of motion is

 \be
 \nab_{\a\bdt}\bar\L^{\bdt}_i=i[W_{ij},\L_{\a}^j]\hspace{4cm}
 \nab_{\adt\b}\L^{\b i}=i[\bar W^{ij},\bar \L_{\adt j}]
 \ee

By differentiating this we find the scalar equation of motion

 \be
 \nab_a\nab^a W_{ij}=\e_{ijkl}[\L_\a{}^k,\L^{\a l}]+2[\bar\L_{\adt
 i},\bar\L^{\adt}_j]+[W_{ik},[W_{jl},\bar W^{kl}]]
 \ee

and the vector equation of motion

 \be \nab_{\adt}{}^{\b}M_{\a \b} = i[\L_{\a}^{i},\bar \L_{\adt i}]
+ \frac{1}{8}([\nab_{\adt \a}W_{ij},\bar{W}^{ij}] -
[W_{ij},\nab_{\adt \a} \bar{W}^{ij}])
 \ee


\section*{Appendix B: Supersymmetry transformations}

The integral invariants we have described in the text are
constructed in terms of the on-shell field strength superfield.
However, we know that if we add in higher-order corrections to the
standard SYM action the supersymmetry transformations will be
modified. As we do not know an off-shell version of this theory,
at least with four off-shell supersymmetries, we could in
principle take this into account by modifying the constraints on
the superspace field strength $F_{AB}$. As we mentioned in the
main text such an approach has been widely discussed in ten
dimensions. In this section we sketch an alternative approach to
the problem which in principle allows one to construct the full
action (up to a given order in $\a'$) starting from on-shell
invariants of the type we have found. The method, a
straightforward generalisation of the BV formalism, allows one
both to complete the action, i.e. to find higher-order terms
induced from lower-order corrections, to find the amended
supersymmetry transformations and to verify that one still has a
closed algebra. Alternatively, one could use the Noether procedure
\cite{Collinucci:2002gd}.

The idea is the following. Starting from an integral invariant one
can work out the corresponding expression as a spacetime integral
of component fields. One then examines the supersymmetry
transformations of the component action constructed by summing the
invariants. Since we have an on-shell theory the supersymmetry
transformations of the zeroth-order theory only close modulo the
field equations and gauge transformations. Moreover, these
transformations are non-linear. To deal with this systematically
one can use the BRST/BV formalism. This requires the introduction
of ghosts for the gauge symmetry and supersymmetry (the latter are
constant because the supersymmetry is rigid). In addition,
anti-fields for both the physical fields and the ghosts must be
introduced. The anti-fields for the fields have ghost number $-1$
while the anti-fields for the ghosts have ghost number $-2$. The
problem is now to construct an extended (spacetime) action $S$
which satisfies the master equation. If we denote all the fields
and the ghosts together by $\f^i$ and the corresponding
anti-fields by $\f^*_i$ we can define an anti-bracket $(A,B)$ of
two functionals by\footnote{In this appendix the summation
convention is understood to include a spacetime integral.}

 \be
 (A,B)={\d A\over \d\f^i}{\d B\over\d\f^*_i}
 +{\d B\over \d\f^i}{\d A\over\d\f^*_i}
 \ee

This can be viewed as an anti-Poisson bracket related to the
anti-symplectic two-form $\d \f^i\wedge \d\f^*_i$. (It is called
anti-symplectic because it is a Grassmann odd two-form.) The
master equation is

 \be
 (S,S)=0
 \ee

We expand $S$ in powers of $\a'$,

 \be
 S=S_0 + S_2 + S_3 +\ldots
 \ee

where $S_0$ is the (extended) classical action, $S_2$ is the
extended action corresponding to the $F^4$ term and so on. The
extended classical action has the form

 \be
 S_0= S_0^0+ S_0^1+ \cO((\f^*)^2)=S_{cl}+  \f^*_i s\f^i + \cO((\f^*)^2)
 \ee

where $s$ denotes the BRST variation of a field or ghost, and
where the superscript counts the number of anti-fields; $S_{cl}$
is the classical action. This implies that the linear terms in
$\f^*$ at each order encode the modified supersymmetry
transformations.

The procedure is now to examine the master equation order by order
in $\a'$. At zeroth order we have $(S_0,S_0)=0$. This equation has
been solved, so we can move on to second order where we find
$(S_0,S_2)=0$. The leading term of this equation is

 \be
 {\d S_{cl}\over \d\f^i}{\d S_2^1\over\d\f^*_i}
 +{\d S^0_2\over \d\f^i}{\d S^1_0\over\d\f^*_i}=0
 \ee

The second term here is the zeroth order variation of the $\a'^2$
action which we know is an on-shell invariant. So this term has
the form $X^i {\d S_{cl}\over\d \f^i}$. Hence we can solve this
equation by taking ${\d S^1_2\over \d \f^*_i}=-X^i$. This
determines the modified supersymmetry transformations to second
order. Assuming we can complete the solution of the master
equation at this order we can move on to third order where the
situation is very similar. However, at fourth order we find

 \be
 2(S_0,S_4) + (S_2,S_2)=0
 \ee

Applying $(S_0,.)$ to this equation and using the Jacobi identity
we find $(S_0,(S_2,S_2))=0$. Assuming that this has a solution of
the form $(S_2,S_2)=2(S_0,Y_4)$, i.e. $(S_2,S_2)$ is
cohomologically trivial, we shall then again have an equation of
the form $(S_0,S'_4)=0$ which we can tackle in the same way as the
lower-order master equation. There is a theoretical possibility
that the terms such as the one we have just been discussing could
be cohomologically non-trivial; this would then represent an
obstruction to a given term being consistent at higher orders. If
this situation were to arise the term in question would presumably
have to be excluded.

We therefore see that the second- and third-order on-shell
invariants do not require any correction terms, although there
will be corrections to the supersymmetry transformations which can
in principle be determined systematically, while from the fourth
order up the lower-order invariants will induce higher-order
corrections. Indeed, as we have seen, there is no independent pure
$F^6$ invariant in the theory and this indicates that it is
induced by the $F^4$ term.


\begin{thebibliography}{99}

 \bm{fradkin}
 E.S Fradkin and A.A. Tseytlin,
 ``Non-linear electrodynamics from quantised strings,''
 Phys.\ Lett.\ B {\bf 163} (1985) 123;
 A. Abouelsaood, C.G. Callan, C.R. Nappi and S.A. Yost,
 ``Open strings in background gauge fields,''
 Nucl.\ Phys.\ B {\bf 280} (1987) 599;
 R.G. Leigh,
 ``Dirac-Born-Infeld action from Dirichlet sigma model,''
 Mod.\ Phys.\ Lett.\ A {\bf 4} (1989) 2767.


\bibitem{Tseytlin:1997cs}
A.~A.~Tseytlin, ``On non-abelian generalisation of the Born-Infeld
action in string  theory,'' Nucl.\ Phys.\ B {\bf 501} (1997) 41
[arXiv:hep-th/9701125].


\bibitem{Kitazawa:xj}
Y.~Kitazawa, ``Effective Lagrangian For Open Superstring From Five
Point Function,'' Nucl.\ Phys.\ B {\bf 289} (1987) 599.


\bibitem{Bergshoeff:2000ik}
E.~A.~Bergshoeff, M.~de Roo and A.~Sevrin, ``Non-Abelian
Born-Infeld and kappa-symmetry,'' J.\ Math.\ Phys.\  {\bf 42}
(2001) 2872 [arXiv:hep-th/0011018].


\bibitem{Bergshoeff:2001dc}
E.~A.~Bergshoeff, A.~Bilal, M.~de Roo and A.~Sevrin,
``Supersymmetric non-abelian Born-Infeld revisited,'' JHEP {\bf
0107}, 029 (2001) [arXiv:hep-th/0105274].


\bibitem{Koerber:2001uu}
P.~Koerber and A.~Sevrin, ``The non-Abelian Born-Infeld action
through order alpha'**3,'' JHEP {\bf 0110} (2001) 003
[arXiv:hep-th/0108169].


\bibitem{DeFosse:2001mk}
L.~De Fosse, P.~Koerber and A.~Sevrin, ``The uniqueness of the
Abelian Born-Infeld action,'' Nucl.\ Phys.\ B {\bf 603} (2001) 413
[arXiv:hep-th/0103015].


\bibitem{Medina:2002nk}
R. Medina, F. T. Brandt, F. R. Machado, ``The open superstring 5-point
amplitude revisited,'' JHEP {\bf 0207} (2002) 071 
[arXiv:hep-th/0208121].


\bibitem{Koerber:2002zb}
P.~Koerber and A.~Sevrin, ``The non-abelian D-brane effective
action through order alpha'**4,'' JHEP {\bf 0210} (2002) 046
[arXiv:hep-th/0208044].

 \bm{wyll}
 N. Wyllard,
 ``Derivative corrections to D-brane actions with constant
 fields,''
 Nucl.\ Phys.\ B {\bf 598} (2001) 247,
 [arXiv:hep-th/0008125].

 \bm{bilal}
 A. Bilal,
 ``Higher derivative corrections to the non-abelian Born-infeld
 action,''
 Nucl.\ Phys.\ B {\bf 618} (2001) 21,
 [arXiv:hep-th/0106062].




\bibitem{Cederwall:1996pv}
M.~Cederwall, A.~von Gussich, B.~E.~Nilsson and A.~Westerberg,
``The Dirichlet super-three-brane in ten-dimensional type IIB
supergravity,'' Nucl.\ Phys.\ B {\bf 490} (1997) 163
[arXiv:hep-th/9610148]: M.~Cederwall, A.~von Gussich,
B.~E.~Nilsson, P.~Sundell and A.~Westerberg, ``The Dirichlet
super-p-branes in ten-dimensional type IIA and IIB supergravity,''
Nucl.\ Phys.\ B {\bf 490} (1997) 179 [arXiv:hep-th/9611159].


\bibitem{Aganagic:1996pe}
M.~Aganagic, C.~Popescu and J.~H.~Schwarz, ``D-brane actions with
local kappa symmetry,'' Phys.\ Lett.\ B {\bf 393} (1997) 311
[arXiv:hep-th/9610249]; ``Gauge-invariant and gauge-fixed D-brane
actions,'' Nucl.\ Phys.\ B {\bf 495} (1997) 99
[arXiv:hep-th/9612080].


\bibitem{Bergshoeff:1996tu}
E.~Bergshoeff and P.~K.~Townsend, ``Super D-branes,'' Nucl.\
Phys.\ B {\bf 490} (1997) 145 [arXiv:hep-th/9611173].

\bibitem{Sorokin:1999jx}
D.~P.~Sorokin, ``Superbranes and superembeddings,'' Phys.\ Rept.\
{\bf 329} (2000) 1 [arXiv:hep-th/9906142].

\bibitem{Howe:1996mx}
P.~S.~Howe and E.~Sezgin, ``Superbranes,'' Phys.\ Lett.\ B {\bf
390} (1997) 133 [arXiv:hep-th/9607227].

 \bm{hrs} P.S. Howe, O. Raetzel and E. Sezgin,
 ``On brane actions and superembeddings,''
 JHEP {\bf 9808} (1998) 011,
 [arXiv:hep-th/9804051].


\bibitem{Howe:1996yn}
P.~S.~Howe and E.~Sezgin, ``D = 11, p = 5,'' Phys.\ Lett.\ B {\bf
394} (1997) 62 [arXiv:hep-th/9611008].

\bibitem{Drummond:2001uj}
J.~M.~Drummond and P.~S.~Howe, ``Codimension zero
superembeddings,'' Class.\ Quant.\ Grav.\  {\bf 18} (2001) 4477
[arXiv:hep-th/0103191].


\bibitem{Howe:2001wc}
P.~S.~Howe and U.~Lindstrom, ``Kappa-symmetric higher derivative
terms in brane actions,'' Class.\ Quant.\ Grav.\  {\bf 19} (2002)
2813 [arXiv:hep-th/0111036].

\bibitem{Drummond:2002kg}
J.~M.~Drummond, P.~S.~Howe and U.~Lindstrom, ``Kappa-symmetric
non-Abelian Born-Infeld actions in three dimensions,'' Class.\
Quant.\ Grav.\  {\bf 19} (2002) 6477 [arXiv:hep-th/0206148].


\bibitem{Sorokin:2001av}
D.~P.~Sorokin, ``Coincident (super)-Dp-branes of codimension
one,'' JHEP {\bf 0108} (2001) 022 [arXiv:hep-th/0106212].

\bibitem{Panda:2003dj}
S.~Panda and D.~Sorokin,
``Supersymmetric and kappa-invariant coincident D0-branes,''
JHEP {\bf 0302} (2003) 055
[arXiv:hep-th/0301065].


\bibitem{Kerstan:2002au}
S.~F.~Kerstan, ``Supersymmetric Born-Infeld from the D9-brane,''
Class.\ Quant.\ Grav.\  {\bf 19} (2002) 4525
[arXiv:hep-th/0204225].

\bibitem{Howe:mf}
P.~S.~Howe, ``Pure Spinors Lines In Superspace And Ten-Dimensional
Supersymmetric Theories,'' Phys.\ Lett.\ B {\bf 258} (1991) 141
[Addendum-ibid.\ B {\bf 259} (1991) 511].


\bibitem{Howe:1991bx}
P.~S.~Howe, ``Pure spinors, function superspaces and supergravity
theories in ten-dimensions and eleven-dimensions,'' Phys.\ Lett.\
B {\bf 273} (1991) 90.

\bibitem{Cederwall:2001dx}
M.~Cederwall, B.~E.~Nilsson and D.~Tsimpis, ``Spinorial cohomology
and maximally supersymmetric theories,'' JHEP {\bf 0202} (2002)
009 [arXiv:hep-th/0110069].


\bibitem{Berkovits:2000nn}
N.~Berkovits, ``Cohomology in the pure spinor formalism for the
superstring,'' JHEP {\bf 0009} (2000) 046 [arXiv:hep-th/0006003].

 \bm{brs}
 E. Bergshoeff, M. Rakowski and E. Sezgin,
 ``Higher-derivative super Yang-Mills theory,''
 Phys.\ Lett.\ B{\bf 185} (1987) 371.


\bibitem{Cederwall:2001td}
M.~Cederwall, B.~E.~Nilsson and D.~Tsimpis, ``D = 10
super-Yang-Mills at O(alpha**2),'' JHEP {\bf 0107} (2001) 042
[arXiv:hep-th/0104236].


\bibitem{Cederwall:2002df}
M.~Cederwall, B.~E.~Nilsson and D.~Tsimpis, ``Spinorial cohomology
of abelian d = 10 super-Yang-Mills at  O(alpha'**3),'' JHEP {\bf
0211}, 023 (2002) [arXiv:hep-th/0205165].

\bibitem{deRoo:2002ap}
M.~de Roo, M.~G.~Eenink, P.~Koerber and A.~Sevrin, ``Testing the
fermionic terms in the non-abelian D-brane effective  action
through order alpha'**3,'' JHEP {\bf 0208} (2002) 011
[arXiv:hep-th/0207015].


\bibitem{Collinucci:2002gd}
A.~Collinucci, M.~de Roo and M.~G.~Eenink, ``Derivative
corrections in 10-dimensional super-Maxwell theory,'' JHEP {\bf
0301} (2003) 039 [arXiv:hep-th/0212012].

\bibitem{Refolli:2001df}
A.~Refolli, A.~Santambrogio, N.~Terzi and D.~Zanon, ``F**5
contributions to the nonabelian Born Infeld action from a
supersymmetric Yang-Mills five-point function,'' Nucl.\ Phys.\ B
{\bf 613} (2001) 64 [Erratum-ibid.\ B {\bf 648} (2003) 453]
[arXiv:hep-th/0105277].

 \bm{grasso}
 D.T. Grasso,
 ``Higher-order contributions to the effective action of N=4 super
 Yang-Mills,''
 [arXiv:hep-th/0210146].

 \bm{kiritsis}
 E. Kiritsis,
 ``Duality and Instantons in string theory,''
 Lectures at the 1999 Trieste Spring School,
 [arXiv:hep-th/9906018].

\bibitem{Howe:1981xy}
P.~S.~Howe, K.~S.~Stelle and P.~K.~Townsend, ``Superactions,''
Nucl.\ Phys.\ B {\bf 191} (1981) 445.

\bibitem{Galperin:1984av}
A.~Galperin, E.~Ivanov, S.~Kalitsyn, V.~Ogievetsky and
E.~Sokatchev, ``Unconstrained N=2 Matter, Yang-Mills And
Supergravity Theories In Harmonic Superspace,'' Class.\ Quant.\
Grav.\  {\bf 1} (1984) 469.

\bibitem{Galperin:uw}
A.~S.~Galperin, E.~A.~Ivanov, V.~I.~Ogievetsky and
E.~S.~Sokatchev, ``Harmonic Superspace,'' Cambridge University
Press (2001).

\bibitem{Hartwell:1994rp}
P.S. Howe and G.G. Hartwell, ``A superspace survey,'' Class.
quantum Grav. {\bf 12} (1995) 1823-1880, G.~G.~Hartwell and
P.~S.~Howe, ``(N, p, q) harmonic superspace,'' Int.\ J.\ Mod.\
Phys.\ A {\bf 10} (1995) 3901 [arXiv:hep-th/9412147].

\bibitem{screp}
M.Flato and C. Fronsdal,  Lett. Math. Phys. {\bf 8} (1984) 159;
V.K. Dobrev and V.B. Petkova, Phys. Lett. {\bf B162} (1985) 127,
Fortschr. Phys. {\bf 35} (1987) 537; B. Binegar, Phys. Rev. {\bf
D34} (1986) 525; B. Morel, A. Sciarrino and P. Sorba, Phys. Lett
{\bf B166} (1986) 69, erratum {\bf B167} (1986) 486.

 \bm{ferrara}
 L. Andrianopoli and S. Ferrara,
 ``K-K excitations on $AdS_5\xz S^5$ as $N=4$ primary
 superfields,''
 Phys.\ Lett.\ B {\bf 430} (1998) 248, [arXiv:hep-th/9803171]'
 ``On long an short $SU(2,2|4)$ multiplets in the AdS/CFT
 correspondence,''
 Lett.\ Math.\ Phys.\ {\bf 48} (1999) 145, [arXiv:hep-th/9812067];
 S. Ferrara and A. Zaffaroni,,
 ``Superconformal field theories, multiplet shortening and the
 $AdS_5/SCFT_4$ correspondence,''
 [arXiv:hep-th/99080163].

 \bm{hw}
 P.S. Howe and P.C. West,
 ``Non-perturbative Green's functions in theories with extended
 supersymmetry,''
 Int.\ J.\ Mod.\ Phys.\ {\bf 10} (1999) 2559,
 [arXiv:hep-th/9509140].



\bibitem{afsz}
L. Adrianopoli, S. Ferrara, E. Sokatchev and B. Zupnik,
``Shortening of primary operators in $N$-extended SCFT and
harmonic superspace analyticity," Adv. Theor. Math. Phys. {\bf 3}
(1999) 1149-1197, [arXiv:hep-th/9912007].

\bibitem{fs}
S. Ferrara and E. Sokatchev, ``Short representations of
$SU(2,2|N)$ and harmonic superspace analyticity,"
 Lett. Math. Phys.
{\bf 52} (2000) 247-262, [arXiv:hep-th/9912168]; ``Superconformal
interpretation of BPS states in AdS geometry," Int. J. Theor.
Phys. {\bf 40} (2001) 935-984, [arXiv:hep-th/0005151].

\bm{hesh2}
 P.~Heslop and P.~S.~Howe,
``On harmonic superspaces and superconformal fields in four
dimensions," Class.\ Quant.\ Grav.\  {\bf 17} (2000) 3743;
[arXiv:hep-th/0005135].

\bibitem{Ryzhov:2001bp}
A.~V.~Ryzhov, ``Quarter BPS operators in N = 4 SYM,'' JHEP {\bf
0111} (2001) 046 [arXiv:hep-th/0109064].


\bibitem{D'Hoker:2003vf}
E.~D'Hoker, P.~Heslop, P.~Howe and A.~V.~Ryzhov, ``Systematics of
quarter BPS operators in N = 4 SYM,'' JHEP {\bf 0304} (2003) 038
[arXiv:hep-th/0301104].


\bibitem{Howe:1981qj}
P.~S.~Howe, K.~S.~Stelle and P.~K.~Townsend, ``Supercurrents,''
Nucl.\ Phys.\ B {\bf 192} (1981) 332.

\bm{bandos} I.A. Bandos, {\sl On the solutions of linear equations
on Lorentz harmonic superspaces}, Theor. Math. Phys. {\bf 76}
(1988) N2 783-792 [169-183].


\bibitem{Ivanov:2001ec}
E.~A.~Ivanov and B.~M.~Zupnik,
``N = 3 supersymmetric Born-Infeld theory,''
Nucl.\ Phys.\ B {\bf 618} (2001) 3
[arXiv:hep-th/0110074].

\bibitem{Howe:1998jw}
P.~S.~Howe, ``On harmonic superspace," Contributed to
International Seminar on Supersymmetries and Quantum Symmetries
(Dedicated to the Memory of Victor I. Ogievetsky), Dubna, Russia,
22-26 Jul 1997. In *Dubna 1997, Supersymmetries and quantum
symmetries* 68-78. [arXiv:hep-th/9812133], ``Aspects of the
$D=6,(2,0)$ tensor multiplet,'' Phys.\ Lett.\ B {\bf 503} (2001)
197,[arXiv:hep-th/0008048].

\bm{Leeming} P.S. Howe and M.I.Leeming, ``Harmonic superspaces in
low dimensions," Class.\ Quant.\ Grav.\  {\bf 11} (1994) 2843
[arXiv:hep-th/9408062].

\bibitem{Ferrara:2000xg}
S.~Ferrara and E.~Sokatchev, ``Representations of (1,0) and (2,0)
superconformal algebras in six  dimensions: Massless and short
superfields,'' Lett.\ Math.\ Phys.\  {\bf 51} (2000) 55
[arXiv:hep-th/0001178].

\bibitem{Ferrara:2000ki}
S.~Ferrara and E.~Sokatchev, ``Conformal primaries of OSp(8/4,R)
and BPS states in AdS(4),'' JHEP {\bf 0005} (2000) 038
[arXiv:hep-th/0003051].

\bibitem{Pasti:1997gx}
P.~Pasti, D.~P.~Sorokin and M.~Tonin, ``Covariant action for a D =
11 five-brane with the chiral field,'' Phys.\ Lett.\ B {\bf 398}
(1997) 41 [arXiv:hep-th/9701037].

 \bm{hklt}
 P.S. Howe, S. Kerstan, U. Lindstr\"om and D. Tsimpis, in
 preparation.

 \bibitem{Kallosh:1980fi}
R.~E.~Kallosh, ``Counterterms In Extended Supergravities,'' Phys.\
Lett.\ B {\bf 99} (1981) 122.

\bibitem{Bern:2000mf}
Z.~Bern, L.~J.~Dixon, D.~C.~Dunbar, A.~K.~Grant, M.~Perelstein and
J.~S.~Rozowsky, ``On perturbative gravity and gauge theory,''
Nucl.\ Phys.\ Proc.\ Suppl.\  {\bf 88} (2000) 194
[arXiv:hep-th/0002078].

\bibitem{Howe:2002ui}
P.~S.~Howe and K.~S.~Stelle, ``Supersymmetry counterterms
revisited,'' Phys.\ Lett.\ B {\bf 554} (2003) 190
[arXiv:hep-th/0211279].

\bibitem{Intriligator:1998ig}
K.~A.~Intriligator, ``Bonus symmetries of N = 4 super-Yang-Mills
correlation functions via  AdS duality,'' Nucl.\ Phys.\ B {\bf
551} (1999) 575 [arXiv:hep-th/9811047].

\bm{sohnius} M.F. Sohnius ``Bianchi Identities for supersymmetric
gauge theories,'' Nucl. Phys. B {\bf 136} (1978) 461.

\end{thebibliography}
\end{document}